\documentclass[11pt,a4paper]{article}
\usepackage{longtable}
\usepackage{amsmath}
\usepackage{amssymb}
\usepackage{graphicx}
\usepackage{bm}% bold math
\usepackage{footmisc}
\usepackage{afterpage}
\usepackage{float}
\usepackage{lscape}
\usepackage{longtable}
\usepackage{float}
\usepackage{slashed}
\usepackage[force]{feynmp-auto}

\newcommand*{\no}{\noindent}
\newcommand*{\bea}{\begin{eqnarray}}
\newcommand*{\eea}{\end{eqnarray}}
\newcommand*{\be}{\begin{equation}}
\newcommand*{\ee}{\end{equation}}
\newcommand*{\pd}{\partial}
\newcommand*{\pdm}{\pd_{\mu}}

\newcommand*{\pref}[1]{(\ref{#1})}

\newcommand*{\mn}{{\mu\nu}}

\newcommand*{\prefr}[2]{(\ref{#1}-\ref{#2})} 
\newcommand*{\nn}{\nonumber}

\newcommand*{\diag}{\mathrm{diag}}

\newcommand{\bma}{\begin{pmatrix}}
\newcommand{\ema}{\end{pmatrix}}

\usepackage[square,comma,sort&compress,numbers]{natbib}

\title{Dyson-Schwinger equations and ${\cal N}=4$ SYM in Landau gauge}

\author{Axel Maas, Stefan Zitz\\
Institute of Physics, NAWI Graz, University of Graz,\\
Universit\"atsplatz 5, A-8010 Graz, Austria}

\begin{document}

\maketitle

\begin{abstract}

${\cal N}=4$ Super Yang-Mills theory is a highly constrained theory, and therefore a valuable tool to test the understanding of less constrained Yang-Mills theories. Our aim is to use it to test our understanding of both the Landau gauge beyond perturbation theory as well as truncations of Dyson-Schwinger equations in ordinary Yang-Mills theories. We derive the corresponding equations within the usual one-loop truncation for the propagators after imposing the Landau gauge. We find a conformal solution in this approximation, which surprisingly resembles many aspects of ordinary Yang-Mills theories. We furthermore discuss which role the Gribov-Singer ambiguity in this context could play, should it exist in this theory.

\end{abstract}

\section{Introduction}

The properties of the gauge-dependent correlation functions of Yang-Mills theories, especially in Landau gauge, have been a long-standing problem, see \cite{Maas:2011se,Alkofer:2000wg,Fischer:2006ub,Binosi:2009qm,Boucaud:2011ug,Vandersickel:2012tg} for reviews. The central problem is that Yang-Mills theory is strongly coupled, and contains long-range correlations. As a consequence, both lattice and continuum methods face severe challenges in calculating the correlation functions. On top of this problem comes that the definition of Landau gauge itself is becoming complicated beyond perturbation theory, due to the Gribov-Singer ambiguity \cite{Gribov:1977wm,Singer:1978dk,Maas:2011se,Vandersickel:2012tg}.

To understand the arising problems better has led to the consideration of simpler theories, especially two-dimensional Yang-Mills theories \cite{Maas:2007uv,Cucchieri:2007rg,Cucchieri:2008fc,Dudal:2008xd,Reinhardt:2008ij,Huber:2012td,Dudal:2012td} where the absence of dynamics reduced the problem to a pure gauge-fixing problem. However, the Gribov-Singer ambiguity appears to be only slightly less severe in this case in terms of numbers of copies \cite{Maas:2015nva,Maas:2011se}. Therefore, it would be useful to find a gauge theory where the problem is even further simplified.

One candidate for this is ${\cal N}=4$ Super-Yang-Mills theory (SYM) \cite{Weinberg:2000cr}. Due to the superconformal symmetry, there is also no conventional dynamics in the theory, though not in the same sense as in two-dimensional Yang-Mills theory. Furthermore, the conformal symmetry should restrict the properties of correlation functions substantially. Finally, some arguments have been made that the Gribov-Singer ambiguity should be less severe in this theory \cite{Capri:2014tta}.

Hence, here we will use ${\cal N}=4$ SYM to test the reliability of the standard truncations employed in Dyson-Schwinger equations (DSEs) \cite{Maas:2011se,Alkofer:2000wg,Fischer:2006ub,Binosi:2009qm,Boucaud:2011ug}. To this end, we derive the truncated Dyson-Schwinger equations in Landau gauge in section \ref{s:dse}. After that we construct a solution respecting the conformal symmetry in section \ref{s:scale}, demonstrating explicitly that it survives the truncation. Finally, in section \ref{s:gribov} we comment on how to test the Gribov-Singer ambiguity in this context. We conclude with some final remarks in section \ref{s:sum}.

\section{Dyson-Schwinger equations}\label{s:dse}

For the formulation of ${\cal N}=4$ SYM we follow \cite{Weinberg:2000cr}. The Lagrangian is given by
\bea
 \mathcal{L} &=& -\frac{1}{2}(D_{\mu}\phi^{ij})_a(D^{\mu}\phi^{ij})_a^{\ast} -\frac{1}{2}\psi^T_{iaL}\epsilon(\slashed D\psi^i_R)_a+\frac{1}{2}\psi^{iT}_{aR}\epsilon(\slashed D\psi_{iL})_a\nn\\
 &&-\frac{1}{4}F_{a\mu\nu}F^{a\mu\nu}-g\sqrt{2}\Re f^{abc}\phi^{ij}_a(\psi^{T}_{ibL}\epsilon\psi_{jcL})-\frac{g^2}{8}|f^{abc}\phi^{ij}_b\phi^{kl}_c|^2\nn\\
 &&+\frac{g^2\theta}{64\pi^2}\varepsilon_{\mu\nu\rho\sigma}F_a^{\mu\nu}F_a^{\rho\sigma}\nn
\eea
\no which exhibits a manifest SU(4) R-symmetry. Note that the auxiliary fields have been integrated out, and thus the supersymmetry is realized non-linearly off-shell.

The scalar fields $\phi^{ij}_a$ are in the adjoint representation of the gauge group and components of an antisymmetric R-tensor obeying the reality constraint
\be
(\phi_a^{ij})^*=\frac{1}{2}\epsilon_{ijkl}\phi^{kl}_a\nn.
\ee
\no The fields $A_\mu^a$ are the usual Yang-Mills gauge fields, which are R-symmetry singlets, and contained in both the usual field strength tensor $F_\mn^a$ and the adjoint covariant derivatives $D_\mu$. They couple with the gauge coupling $g$. The topological term with vacuum angle $\theta$ will be set to zero, $\theta=0$. Note that such a term would not appear in the DSEs, since it is a boundary term. A non-zero value of $\theta$ would rather surface as a boundary condition to the solutions of the DSEs \cite{Guralnik:2007rx}. Finally, there is a quadruplet of Majorana fermions $\Psi$, where $L$ and $R$ indicate left-handed and right-handed, respectively, and the matrix $\epsilon=\diag(i\sigma_2,i\sigma_2)$ can be used to relate both
\be
\psi^{a*}_{iL}=-\beta\epsilon\psi^{ia}_R\nn
\ee
\no with the $4\times 4$ matrix
\be
\beta=\bma 0 & 1 \cr 1 & 0 \ema\nn.
\ee
\no In the following, correlation functions will be considered for simplicity of complex scalar and fermion fields, though when deriving the DSEs the relations between them have been duly taken into account.

It is furthermore useful to define
\be
\delta_{abcd}=\delta_{ac}\delta_{bd}-\delta_{ad}\delta_{bc}=\frac{1}{2}\varepsilon_{abij}\varepsilon_{ijcd}\nn,
\ee
\no as this combination will appear repeatedly.

The Lagrangian, as it stands, is not yet gauge-fixed. As our aim is to compare to the ordinary Yang-Mills theory, we need to fix to a gauge also realizable there. It is hence not possible to chose a manifestly supersymmetric gauge. Rather, we will use the Landau gauge
\be
\pdm A^\mu_a=0\nn,
\ee
\no which is the gauge best studied in ordinary Yang-Mills theory \cite{Maas:2011se,Alkofer:2000wg,Fischer:2006ub,Binosi:2009qm,Boucaud:2011ug,Vandersickel:2012tg}. Thus, this gauge condition hides supersymmetry, and therefore the gauge-dependent correlation functions will not be manifestly supersymmetric\footnote{In particular, this implies that the supersymmetric WTIs will become modified, as only a diagonal subgroup of the BRST symmetry and the supersymmetry remains unbroken. This will link the supersymmetry WTIs to the STIs, which, however, involve higher order correlation functions, making them of very limited use in DSE studies \cite{Maas:2011se,Fischer:2008uz}.}. Especially, this also implies that renormalization is in general necessary \cite{Storey:1981ee}. However, since the Landau gauge condition is conformally invariant, we assume that conformal symmetry remains manifest. Our results in section \ref{s:scale} turn indeed out to be conformally invariant, i.\ e.\ pure power laws.

The gauge-fixing is performed in the same way as in ordinary Yang-Mills theory, and requires the introduction of a ghost Lagrangian
\be
\mathcal{L}_{Gh} = \overline{c^{a}}\partial_{\mu}D^{ab}_{\mu}c^{b}\nn
\ee
\no with a ghost and antighost fields $c$ and $\overline{c}$ in the adjoint representation, and which are singlets under the R-symmetry. Of course, there is the Gribov-Singer ambiguity, to which we will only return in section \ref{s:gribov}, as this will not modify the following.

To derive the DSEs, we Wick rotate to Euclidean space-time, and follow the standard procedures to derive the DSEs \cite{Maas:2011se,Alkofer:2000wg,Fischer:2006ub,Binosi:2009qm,Boucaud:2011ug}. However, our aim here is to investigate the implication of truncation schemes on the solutions of DSEs. Thus, we truncate the propagator DSEs at the non-perturbative one-loop level, i.\ e.\ keeping only the equations for the propagators, but looking for self-consistent solutions for them. This requires to model the vertices. The choice of models will be discussed in section \ref{s:scale}. This type of truncation, including the gauge-fixing dynamics, was first introduced in \cite{vonSmekal:1997is}. More sophisticated truncations show quantitatively better results \cite{Huber:2012kd,Eichmann:2014xya,Blum:2014gna,Aguilar:2013xqa,Fister:2013bh}, but the qualitative features appear to be captured already correctly by this truncation \cite{Alkofer:2000wg,Fischer:2006ub,Binosi:2009qm,Maas:2011se,Boucaud:2011ug,Fischer:2008uz}.

\begin{figure}
%Ghost equation
\be 
 \parbox{10mm}{\begin{fmffile}{ghost0}
	\begin{fmfgraph*}(40,40)
	\fmfleft{in} \fmfright{out}
	\fmf{dots}{in,1} \fmf{dots}{1,out}
	\fmfdot{1}
	\end{fmfgraph*}
\end{fmffile}}
	^{-1}\qquad =\qquad
\parbox{10mm}{\begin{fmffile}{ghost1}
	\begin{fmfgraph*}(40,40)
	\fmfleft{in} \fmfright{out}
	\fmf{dots}{in,out}
	\end{fmfgraph*}
\end{fmffile}}
	^{-1}\qquad+\qquad
\parbox{10mm}{\begin{fmffile}{ghost2}
	\begin{fmfgraph*}(40,30)
	\fmfleft{in} \fmfright{out} \fmftop{on} \fmfbottom{down}
	\fmf{dots}{in,v1} \fmf{dots}{v2,out}
	\fmf{dots,tension=1/3}{v1,v,v2}
	\fmf{curly,left,tension=0}{v1,v2}
	\fmfdot{v,on}
	\end{fmfgraph*}
	\end{fmffile}}\nn\\
\ee
\bea
\parbox{10mm}{\begin{fmffile}{glu1}
	\begin{fmfgraph*}(40,40)
	\fmfleft{in} \fmfright{out}
	\fmf{curly}{in,v1,out}
	\fmfdot{v1}
	\end{fmfgraph*}
\end{fmffile}}
	^{-1}\quad =\quad&
\parbox{10mm}{\begin{fmffile}{glu2}
	\begin{fmfgraph*}(40,42)
	\fmfleft{in} \fmfright{out}
	\fmf{curly}{in,out}
	\end{fmfgraph*}
\end{fmffile}}
	^{-1}\quad+\quad
\parbox{10mm}{\begin{fmffile}{glu3}
	\begin{fmfgraph*}(40,22)
	\fmfleft{in} \fmfright{out} \fmftop{on} \fmfbottom{down}
	\fmf{curly}{in,v1} \fmf{curly}{v2,out}
	\fmf{phantom}{on,down}
	\fmf{dots,left,tension=0.2,tag=1}{v1,v2}
	\fmf{dots,left,tension=0.2,tag=2}{v2,v1}
	\fmfdot{on,down}
	\end{fmfgraph*}
	\end{fmffile}}
		\qquad+\quad
\parbox{10mm}{\begin{fmffile}{glu4}
	\begin{fmfgraph*}(40,22)
	\fmfleft{in} \fmfright{out} \fmftop{on} \fmfbottom{down}
	\fmf{curly}{in,v1} \fmf{curly}{v2,out}
	\fmf{curly,left,tension=0.2,tag=1}{v1,v2}
	\fmf{curly,left,tension=0.2,tag=2}{v2,v1}
	\fmf{phantom}{on,down}
	\fmfdot{on,down}
	\end{fmfgraph*}
	\end{fmffile}}
		\qquad+\quad
		\parbox{10mm}{\begin{fmffile}{glu7}
	\begin{fmfgraph*}(40,22)
	\fmfleft{in} \fmfright{out} \fmftop{on} \fmfbottom{down}
	\fmf{curly}{in,v1} \fmf{curly}{v2,out}
	\fmf{dashes,left,tension=0.2,tag=1}{v1,v2}
	\fmf{dashes,left,tension=0.2,tag=2}{v2,v1}
	\fmfdot{on,down}
	\end{fmfgraph*}
	\end{fmffile}}\nn\\
		&\qquad+\quad
\parbox{10mm}{\begin{fmffile}{glu6}
	\begin{fmfgraph*}(40,22)
	\fmfleft{in} \fmfright{out} \fmftop{on} \fmfbottom{down}
	\fmf{curly}{in,v1} \fmf{curly}{v2,out}
	\fmf{plain,left,tension=0.2,tag=1}{v1,v2}
	\fmf{plain,left,tension=0.2,tag=2}{v2,v1}
	\fmfdot{on,down}
	\end{fmfgraph*}
	\end{fmffile}}
		\qquad+\quad
\parbox{10mm}{\begin{fmffile}{glu5}
	\begin{fmfgraph*}(40,50)
	\fmfleft{in} \fmfright{out} \fmftop{on}
	\fmf{curly}{in,v1} \fmf{curly}{v1,out}
	\fmf{curly,right,tension=0.8,tag=1}{v1,v1}
	\fmfdot{on}
	\end{fmfgraph*}
	\end{fmffile}}
		\qquad+\quad
\parbox{10mm}{\begin{fmffile}{glu8}
	\begin{fmfgraph*}(40,40)
	\fmfleft{in} \fmfright{out} \fmftop{on}
	\fmf{curly}{in,v1} \fmf{curly}{v1,out}
	\fmf{dashes,right,tension=0.7,tag=1}{v1,v1}
	\fmfdot{on}
	\end{fmfgraph*}
	\end{fmffile}}\nn
\eea
\bea
\parbox{10mm}{\begin{fmffile}{ferm1_0}
	\begin{fmfgraph*}(40,40)
	\fmfleft{in} \fmfright{out}
	\fmf{plain}{in,v,out}
	\fmfdot{v}
	\end{fmfgraph*}
\end{fmffile}}
	^{-1}\qquad =\qquad
\parbox{10mm}{\begin{fmffile}{ferm1_1}
	\begin{fmfgraph*}(40,40)
	\fmfleft{in} \fmfright{out}
	\fmf{plain}{in,out}
	\end{fmfgraph*}
\end{fmffile}}
	^{-1}\qquad+\qquad
\parbox{10mm}{\begin{fmffile}{ferm1_2}
	\begin{fmfgraph*}(40,30)
	\fmfleft{in} \fmfright{out} \fmftop{on} \fmfbottom{down}
	\fmf{plain}{in,v1} \fmf{plain}{v2,out}
	\fmf{plain,tension=1/3}{v1,i,v2}
	\fmf{curly,left,tension=0}{v1,v2}
	\fmfdot{i,on}
	\end{fmfgraph*}
	\end{fmffile}}
		\qquad+\qquad
\parbox{10mm}{\begin{fmffile}{ferm1_3}
	\begin{fmfgraph*}(40,30)
	\fmfleft{in} \fmfright{out} \fmftop{on} \fmfbottom{down}
	\fmf{plain}{in,v1} \fmf{plain}{v2,out}
	\fmf{plain,tension=1/3}{v1,i,v2}
	\fmf{dashes,left,tension=0}{v1,v2}
	\fmfdot{on,i}
	\end{fmfgraph*}
	\end{fmffile}}\nn
\eea
\bea
\parbox{10mm}{\begin{fmffile}{sc0}
	\begin{fmfgraph*}(40,40)
	\fmfleft{in} \fmfright{out}
	\fmf{dashes}{in,v,out}
	\fmfdot{v}
	\end{fmfgraph*}
\end{fmffile}}
	^{-1}\qquad =\qquad &
\parbox{10mm}{\begin{fmffile}{sc1}
	\begin{fmfgraph*}(40,40)
	\fmfleft{in} \fmfright{out}
	\fmf{dashes}{in,out}
	\end{fmfgraph*}
\end{fmffile}}
	^{-1}\qquad+\qquad
\parbox{10mm}{\begin{fmffile}{sc2}
	\begin{fmfgraph*}(40,30)
	\fmfleft{in} \fmfright{out} \fmftop{on} \fmfbottom{down}
	\fmf{dashes}{in,v1} \fmf{dashes}{v2,out}
	\fmf{dashes,tension=1/3}{v1,i,v2}
	\fmf{curly,left,tension=0}{v1,v2}
	\fmfdot{i,on}
	\end{fmfgraph*}
	\end{fmffile}}
		\qquad+\qquad
\parbox{10mm}{\begin{fmffile}{sc3}
\begin{fmfgraph*}(40,22)
	\fmfleft{in} \fmfright{out} \fmftop{on} \fmfbottom{down}
	\fmf{dashes}{in,v1} \fmf{dashes}{v2,out}
	\fmf{plain,left,tension=0.2,tag=1}{v1,v2}
	\fmf{plain,left,tension=0.2,tag=2}{v2,v1}
	\fmfdot{on,down}
	\end{fmfgraph*}
	\end{fmffile}}\nn\\
		&\qquad+\qquad
\parbox{10mm}{\begin{fmffile}{sc4}
	\begin{fmfgraph*}(40,50)
	\fmfleft{in} \fmfright{out} \fmftop{on}
	\fmf{dashes}{in,v1} \fmf{dashes}{v1,out}
	\fmf{curly,right,tension=0.8,tag=1}{v1,v1}
	\fmfdot{on}
	\end{fmfgraph*}
	\end{fmffile}}
		\qquad+\quad
\parbox{10mm}{\begin{fmffile}{sc5}
	\begin{fmfgraph*}(40,40)
	\fmfleft{in} \fmfright{out} \fmftop{on}
	\fmf{dashes}{in,v1} \fmf{dashes}{v1,out}
	\fmf{dashes,right,tension=0.7,tag=1}{v1,v1}
	\fmfdot{on}
	\end{fmfgraph*}
	\end{fmffile}}\nn
\eea
\caption{\label{dses}The DSEs in the employed truncation scheme. Objects with a large dot are full, while all other quantities are either bare or modeled, see text for details. Dotted lines are ghosts, curly lines are gluons, dashed lines are scalars, and solid lines are fermions.}
\end{figure}

The resulting four DSEs for the ghost, the gluon, the fermion, and the scalar are graphically shown in figure \ref{dses}. These equations will now be discussed in turn.

The ghost equation is given by
\bea
 D_G^{ab}(p)^{-1} &=& -\delta^{ab}p^2+ \int\frac{d^4q}{(2\pi)^4}\Gamma^{tl,c\overline{c}A;dae}_{\mu}(-q,p,q-p)\times\nn\\
 &&\times D^{ef}_{\mu\nu}(p-q)D_G^{dg}(q)\Gamma^{c\overline{c}A;bgf}_{\nu}(-p,q,p-q)\nn
\eea
\no which is identical to the one of Yang-Mills theory \cite{Alkofer:2000wg}. The ghost and gluon propagators are described by single scalar dressing functions
\bea
D_G^{ab}(p)&=&-\frac{\delta^{ab}}{p^2}G(p)\nn\\
D_\mu^{ab}&=&\delta^{ab}\left(\delta_\mn-\frac{p_\mu p_\nu}{p^2}\right)\frac{Z(p)}{p^2}\nn
\eea
\no and 
\be
 \Gamma^{tl,c\overline{c}A;dae}_{\mu}(p,q,k) = igf^{dae}q_{\mu}\delta(p+q+k)\nn
\ee
\no is the tree-level\footnote{We have checked the tree-level results against \cite{Tarasov:2013zv}, though note that there different conventions are used than here.} ghost-gluon vertex, while the quantity without the superscript $tl$ is the full ghost-gluon vertex. 

The gluon equation is given by
\bea
&& D_{\mu\nu}^{ab}(p)^{-1}=\delta^{ab}(\delta_{\mu\nu}p^2-p_{\mu}p_{\nu})\nonumber \\
&&-\int\frac{d^dq}{(2\pi)^d}\Gamma^{tl;Ac\overline{c};dca}_{\mu}(-p-q,q,p)D^{cf}_G(q)D^{de}_G(p+q)\Gamma^{Ac\overline{c};feb}_{\nu}(-q,p+q,-p) \nonumber \\
&&+\frac{1}{2}\int\frac{d^dq}{(2\pi)^d}\Gamma^{tl;A^3;acd}_{\mu\sigma\chi}(p,q-p,-p)D^{cf}_{\sigma\omega}(q)D^{de}_{\chi\lambda}(p-q)\Gamma^{A^3;bfe}_{\nu\omega\lambda}(-p,q,p-q)\nonumber \\
&&+\frac{1}{2}\int\frac{d^dq}{(2\pi)^d}\Gamma^{tl;A\phi^{\ast}\phi;acd}_{\mu;uvrs}(p,q-p,-p)D^{de}_{ijmn}(q)D^{cf}_{ijkl}(p-q)\Gamma^{A\phi^{\ast}\phi;bfe}_{\nu;mnkl}(-p,q,p-q)\nonumber\\
&&-\frac{1}{2}\int\frac{d^dq}{(2\pi)^d}\Gamma^{tl;A\overline{\psi}\psi;acd}_{\mu;uv}(p,q-p,-p)D^{cf}_{ij}(q)D^{de}_{ik}(p-q)\Gamma^{A\overline{\psi}\psi;bfe}_{\nu;jk}(-p,q,p-q)\nonumber \\
&&+\frac{1}{2}\int\frac{d^dq}{(2\pi)^d}\Gamma^{tl;A^4;abcd}_{\mu\nu\rho\sigma}(p,-p,q,-q)D^{cd}_{\rho\sigma}(q)\nonumber \\
&&-\frac{1}{2}\int\frac{d^dq}{(2\pi)^d}\Gamma^{tl;A^2\phi^{\ast}\phi;abdf}_{\mu\nu;ijkl}(p,-p,q,-q)D^{df}_{mnmn}(q)\nn
\eea
\no where in addition the scalar and fermion propagators
\bea
D_{ijkl}^{ab}(p^2)&=&\delta_{ijkl}\delta^{ab}\frac{S(p^2)}{p^2}\nn\\
D_{ij\alpha\beta}^{ab}(p^2)&=&-i\delta_{ij}\delta^{ab}\slashed p_{\alpha\beta}\frac{F(p^2)}{p^2}\nn
\eea
\no appear as well as the tree-level vertices
\bea
\Gamma^{tl;A^3;acd}_{\mu\nu\rho}(p,q,k)&=& -igf^{acd}((q-k)_{\mu}\delta_{\nu\rho}+(k-p)_{\nu}\delta_{\mu\rho}+(p-q)_{\rho}\delta_{\mu\nu})\times\nn\\
&&\times \delta(p+q+k)\nn\\
\Gamma^{tl;A\phi^{\ast}\phi;abc}_{\mu;ijkl}(p,q,k)&=&igf^{abc}(q-k)_{\mu}\delta_{ijkl}\delta(p+q+k)\nn\\
\Gamma^{tl;A\overline{\psi}\psi;abc}_{\mu;ij}(p,q,k)&=&-gf^{abc}\gamma_{\mu}\delta_{ij}\delta(p+q+k)\nn\\
\Gamma^{tl;A^4;abcd}_{\mu\nu\rho\sigma}(p,q,k,l)&=&g^2\left(f^{eab}f^{ecd}(\delta_{\mu\sigma}\delta_{\nu\rho}-\delta_{\mu\rho}\delta_{\nu\sigma})\right.\nn\\
&&\left.+f^{gac}f^{gbd}(\delta_{\mu\nu}\delta_{\sigma\rho}-\delta_{\mu\rho}\delta_{\nu\sigma})\right.\nn\\
&&\left.+f^{gad}f^{gbc}(\delta_{\mu\nu}\delta_{\sigma\rho}-\delta_{\mu\sigma}\delta_{\nu\rho})\right)\delta(p+q+k+l)\nn\\
\Gamma^{tl;A^2\phi^{\ast}\phi;abdf}_{\mu\nu;ijkl}(p,q,k,l)&=&g^2\delta_{\mu\nu}\delta_{ijkl}(f^{abc}f^{adf}+f^{abf}f^{adc}))\delta(p+q+k+l)\nn.
\eea
\no The part only involving the ghosts and gluons is again identical to Yang-Mills theory \cite{Alkofer:2000wg}.

The DSE for the fermion, suppressing Dirac indices, is given by
\bea
&&D^{ab}_{ij}(p)^{-1}= -i\delta_{ij}\delta^{ab}\slashed p \nn\\
&&- \int\frac{d^dq}{(2\pi)^d}\Gamma^{tl;\overline{\psi}\psi A;dae}_{\mu;ik}(-q,p,q-p)D^{ef}_{\mu\nu}(p-q)D^{dg}_{kl}(q)\Gamma^{\overline{\psi}\psi A;bgf}_{\nu;jl}(-p,q,p-q)\nn\\
&&+ \int\frac{d^dq}{(2\pi)^d}\Gamma^{tl;\overline{\psi}\psi\phi^{\ast};dae}_{mn;ik}(-q,p,q-p)D^{ef}_{mnrs}(p-q)D^{dg}_{il}(q)\Gamma^{\overline{\psi}\psi A;bgf}_{rs;jl}(-p,q,p-q)\nn
\eea
\no where only one additional tree-level vertex appears
\bea
\Gamma^{tl;\overline{\psi}\psi\phi^{\ast};abc}_{ij;kl}(p,q,k) = g\frac{\sqrt{2}}{2}f^{abc}(\delta_{ijkl}(1+\gamma_5)-\varepsilon_{ijkl}(1-\gamma_5))\delta(p+q+k)\nn.
\eea
\no This equation is substantially different from the ordinary equation for quarks \cite{Alkofer:2000wg,Fischer:2006ub}, due to the additional Yukawa interaction.

The equation for the scalar is finally
\bea
&&D^{ab}_{ijkl}(p)=\quad \delta_{ijkl}\delta^{ab} p^2\nn \\
&&+ \int\frac{d^dq}{(2\pi)^d}\Gamma^{tl;\phi^{\ast}\phi A;ead}_{\mu;ijmn}(-p-q,p,q)D^{dg}_{\mu\nu}(p+q)D^{fc}_{mnuv}(q)\Gamma^{\phi^{\ast}\phi A;gbf}_{\nu;kluv}(p+q,-p,-q)\nn\\
&&- \int\frac{d^dq}{(2\pi)^d}\Gamma^{tl;\phi^{\ast}\psi\overline{\psi};dca}_{ij;mn}(-p-q,q,p)D^{cf}_{mu}(p+q)D^{de}_{nv}(q)\Gamma^{\phi^{\ast}\psi\overline{\psi};feb}_{kl;uv}(-q,p+q,-p)\nn\\
&&+\int\frac{d^dq}{(2\pi)^d}\Gamma^{tl;\phi^{\ast}\phi A^2;abef}_{\mu\nu;ijkl}(-p,p,-q,q)D^{ef}_{\mu\nu}(q)\nn\\
&&-\frac{1}{4}\int\frac{d^dq}{(2\pi)^d}\Gamma^{tl;\phi^{\ast}\phi\phi^{\ast}\phi;abef}_{ijklmnop}(-p,p,-q,q)D^{ef}_{mnop}(q)\nn,
\eea
\no with the four-scalar tree-level vertex
\bea
\Gamma^{tl;\phi^{\ast}\phi\phi^{\ast}\phi;abcd}_{ijklpqrs}(p,q,k,l)&=&\frac{g^2}{2}\left((f^{ead}f^{abc}+f^{eac}f^{ebd})\frac{1}{2}\varepsilon_{ijkl}\varepsilon_{pqrs}\right.\nn\\
&&+(f^{eab}f^{edc}+f^{eac}f^{edb})\delta_{ijpq}\delta_{klrs}\nn\\
&&\left.+(f^{eab}f^{ecd}+f^{ead}f^{ecb})\delta_{klpq}\delta_{ijrs}\right)\delta(p+q+k+l)\nn.
\eea

\section{A scaling solution}\label{s:scale}

\subsection{Ansatz}

As noted, we do not expect that the conformal symmetry is broken, as it is not explicitly broken by the gauge condition. Hence, we expect that a conformal solution should exist, i.\ e.\ a scaling solution
\bea
G(x)&=&ax^{\kappa_1}\nn\\
Z(x)&=&bx^{\kappa_2}\nn\\
F(x)&=&fx^{\kappa_3}\nn\\
S(x)&=&sx^{\kappa_4}\nn,
\eea
\no where $x=p^2$.

\subsection{Tree-level vertices}

The first thing to note is that the tadpoles can be computed, since only the tree-level vertices are involved. In both the scalar and the gluon equation they cancel each other exactly if $b=3s$. Since if the tadpoles would otherwise lead to divergences, which would require a gauge-non-invariant mass counter-term for the gluon \cite{Maas:2011se}, this relation is imposed, removing the unknown $b$ from the system.

To make progress requires to make ans\"atze for the remaining vertices. The first choice are using the tree-level vertices of section \ref{s:dse}. This yields the reduced equations
\bea
\frac{1}{G(x)}&=&1 - C_Ag^2\int\frac{d^4q}{(2\pi)^4}K(x,y,z)G(y)Z(z)\nn\\
\frac{1}{Z(x)}&=& 1 + C_Ag^2\int\frac{d^4q}{(2\pi)^4}\left(M(x,y,z)G(y)G(z)+N(x,y,z)Z(y)Z(z)\right)\nn\\
&&+ C_Ag^2\int\frac{d^4q}{(2\pi)^4}\left(Q(x,y,z)S(y)S(z) + R(x,y,z)F(y)F(z)\right)\nn\\
\frac{1}{F(x)}&=&1 - C_Ag^2\int\frac{d^4q}{(2\pi)^4}\left(L(x,y,z)F(y)Z(z) +H(x,y,z)F(y)S(z)\right)\nn\\
\frac{1}{S(x)}&=&1 - C_Ag^2\int\frac{d^4q}{(2\pi)^4}\left(D(x,y,z)S(y)Z(z) +B(x,y,z)S(y)F(z)\right)\nn.
\eea
\no Herein $C_A$ is the adjoint Casimir of the gauge group.

The integral kernels are given by
\bea
K(x,y,z)&=&-\frac{x^2+(y-z)^2-2x(y+z)}{4xyz^2}\nn\\
M(x,y,z)&=&-\frac{x^2+(y-z)^2-2x(y+z)}{12x^2yz}\nn\\
N(x,y,z)&=&-\frac{1}{24x^2y^2z^2}(x^4+8x^3(y+z)+(y-z)^2(y^2+10yz+z^2)\nn\\
&&-2x^2(9y^2+16yz+9z^2)+8x(y^3-4y^2z-4yz^2+z^3))\nn\\
Q(x,y,z)&=&-\frac{16}{x^2yz}(x^2+(y-z)^2-2x(y+z))\nn\\
R(x,y,z)&=&-\frac{4}{xyz}(-x+y+z)\nn\\
L(x,y,z)&=&\frac{24}{xyz}(x+y-z)\nn\\
H(x,y,z)&=&\frac{192}{xyz}(x+y-z)\nn\\
D(x,y,z)&=&\frac{96}{xyz^2}(x^2+(y-z)^2-2x(y+z))\nn\\
B(x,y,z)&=&-\frac{96}{xyz}(-x+y+z)\nn
\eea
\no where $x=p^2$ is the external momentum squared, $y=q^2$ is the loop momentum squared, and $z=(p-q)^2$. Since the $R$-symmetry does not alter the momentum dependence, those not involving the fermions agree with the ones of the pure Yang-Mills-scalar theory in the same truncation \cite{Maas:2004se}.

Following \cite{vonSmekal:1997is,Alkofer:2000wg,Fischer:2006ub,Maas:2011se}, the resulting integrals can be calculated using the formula \cite{Peskin:1995ev} 
\be
\int\frac{d^dq}{(2\pi)^d}q^{2\alpha} (q-p)^{2\beta}=\frac{1}{(4\pi)^{\frac{d}{2}}}\frac{\Gamma(-\alpha-\beta-\frac{d}{2})\Gamma(\frac{d}{2}+\alpha)\Gamma(\frac{d}{2}+\beta)}{\Gamma(d+\alpha+\beta)\Gamma(-\alpha)\Gamma(-\beta)}y^{2\left(\frac{d}{2}+\alpha+\beta\right)}.\label{solver}
\ee
\no This formula is actually only correct if the integrals are finite. Otherwise, its usage defines regulated versions of the integrals. In general, the integrals are not finite for arbitrary exponents $\kappa_i$ but can have divergences. As noted above, this is actually expected, due to the choice of Landau gauge. We employ here a renormalization scheme, in which all counter-terms $\delta Z_i$ are chosen as
\be
\delta Z_i=-1+f(\Lambda)\nn
\ee
\no where the function $f(\Lambda)$ is a divergent quantity such that the integrals evaluate exactly to the form \pref{solver}. The addition of the $-1$ is a possible finite shift of the renormalization constants, and is used to cancel all tree-level terms in the following. Note that with the final results many of the integrals are actually finite, and thus the functions $f$ vanish in these cases, leaving only the finite renormalization. Furthermore, no mass counter-terms are necessary, as no divergences requiring them appear.

The resulting set of equations then takes the rather simple form
\bea
\frac{1}{a}x^{-\kappa_1}&=&g_1\times x^{\kappa_1+\kappa_2}\label{eq1}\\
\frac{1}{3s}x^{-\kappa_2}&=&z_1\times x^{2\kappa_1} + z_2\times x^{2\kappa_2} + z_3\times x^{2\kappa_3} + z_4\times x^{2\kappa_4}\label{eq2}\\
\frac{1}{f}x^{-\kappa_3}&=&f_1\times x^{\kappa_2+\kappa_3} + f_2\times x^{\kappa_4+\kappa_3}\label{eq3}\\
\frac{1}{s}x^{-\kappa_4}&=&s_1\times x^{\kappa_2+\kappa_4} + s_2\times x^{2\kappa_3}\label{eq4},
\eea
\no where the precise values of the prefactors $g_1$, $z_i$, $f_i$, and $s_i$ are not relevant for now.

The deceisive difference compared to similar calculations in ordinary Yang-Mills theory \cite{Fischer:2006vf,Fischer:2009tn,Huber:2007kc} is that here a solution is required for all momenta, while in the Yang-Mills case only a solution in the infrared was searched for. As a consequence, it is not possible to consider any of the terms as sub-leading, and in fact a solution is needed which solves all the equations \prefr{eq1}{eq4} simultaneously, and completely for all momenta.

The ghost equation is the logical starting point, as it has only a single term. For any consistent solution, this requires
\be
\kappa_2 = -2\kappa_1\label{shirel},
\ee
\no which is incidentally the same relation as found in the Yang-Mills case \cite{vonSmekal:1997is}. However, this relation ensures that the running coupling, which can be defined in this scheme analogously to the miniMOM scheme in Yang-Mills theory \cite{vonSmekal:1997is,vonSmekal:1997vx,vonSmekal:2009ae,Alkofer:2004it}, as
\be
\alpha(p^2)=\alpha(\mu^2)G^2(p)Z(p)\nn,
\ee
\no is then necessarily momentum independent. Thus, any solution satisfying \pref{shirel} indeed exhibits a conformal behavior in this coupling. In particular, this implies that the present truncation preserves the conformal nature of the theory so far\footnote{Note that if the tree-level term would have not been absorbed in the renormalization process, it would not be possible to find a scaling solution to equation \pref{eq1}, as this would only be possible then for $\kappa_1=\kappa_2=0$, which then always leads to inconsistencies in the other equations, at least in the present truncation.}.

In the equation for the fermion, it is only possible to find a solution, if all three terms show the same scaling. This requires
\be
\kappa_4 = -2\kappa_3\label{shirel2}
\ee
\no thus linking in a very similar way to the ghost and gluon relation \pref{shirel} the exponents of the fermion and the scalar.

However, this leads to a problem in the scalar equation \pref{eq4}, as now here all three terms cannot cancel. Solving the scalar equation first would lead to a similar problem in the fermion equation. Thus, the current truncation does not yield a result.

\subsection{An improved truncation}

There are a multitude of possibilities to do so. Arguably the simplest possibility is by introducing a dressing for the scalar-gluon vertex, since the scalar-gluon vertex can have only a single tensor structure, for the same reason the ghost-gluon vertex has only a single tensor structure in Landau gauge, and because the gluon has a trivial R structure. The situation is more complicated for the fermion-scalar or fermion-gluon vertex.

Thus, the vertex can receive at most a single, Bose-symmetric multiplicative dressing. The simplest conformal possibility is\footnote{Some alternative possibilities are discussed in \cite{Lerche:2002ep} for the case of the ghost-gluon vertex in ordinary Yang-Mills theory.}
\be
\Gamma^{\phi^{\ast}\phi A}(p^2,q^2,k^2)=Sp^{2\alpha_1}q^{2\alpha_1}k^{2\alpha_2}\label{sgdressing}.
\ee
\no This multiplies the corresponding diagrams with a further constant $S$. Note that this is not necessarily the actual behavior of the scalar-gluon vertex, but merely designed to make the system solvable. Including, e.\ g.\ also the two-loop terms or other vertex dressings would potentially lead to a different dressing function.

Due to \pref{solver}, a dressing like \pref{sgdressing} translates into addition in the exponents. The modified result can thus be read off immediately. Requiring
\be
2\alpha_1+\alpha_2=6\kappa_3\nn
\ee
\no then solves simultaneously the fermion and scalar equations, provided the conditions \pref{shirel}, \pref{shirel2}, and 
\be
\kappa_1=\kappa_3\equiv-\kappa\nn,
\ee
\no hold. Then, there is just one single $\kappa$ left.

This ansatz immediately solves also the gluon equation \pref{eq2}, except for the gluon loop. Its remaining exponent is $-4\kappa$, and is thus not compatible with any of the other terms, which all have the exponent $-2\kappa$. Thus, also the three-gluon vertex requires dressing. However, due to the Bose symmetry, any dressing ansatz of the type \pref{sgdressing} can have only a single exponent, $\alpha$. Additionally, there can be four different transverse tensor structures.

To have the same momentum structure as the other terms in the gluon equation requires this single exponent to satisfy
\be
\alpha=2\kappa\nn.
\ee
\no But this value turns out to be pathological, as then the prefactor of $\Gamma$-functions in \pref{solver} actually diverges for all possible tensor structures. The resolution we choose here is to force the prefactors of the different tensor structures to such values that the gluon loop cancels itself for any value of the exponents different from $2\kappa$. Then, this term drops out. This completes our truncation. This is possible for an infinite number of different values of the prefactors, as only their relative size needs to be fixed. In fact, it is even possible to cancel either pairwise two tensor structures, include only three tensor structures, or require all four tensor structures to cancel.

It is interesting to note that this truncation also ensures, for every value of the exponent $\kappa$, that the running couplings derived from the bare fermion-scalar and fermion-gluon vertices \cite{Alkofer:2004it}
\bea
&\alpha_1 F^2(x)S(x)=\alpha_1 f^2s x^{-2\kappa+2\kappa}&\nn\\
&\alpha_2 F^2(x)Z(x)=\alpha_2 3f^2s x^{-2\kappa+2\kappa}&\nn
\eea
\no are then also constant. This is also true, at the symmetric point, for the scalar gluon vertex, since for it follows \cite{Alkofer:2004it}
\be
\alpha_3 x^{6\kappa}S(x)^2Z(x)=\alpha_3 3s^3x^{6\kappa-6\kappa}\nn.
\ee
\no For the three-gluon vertex, this is actually not possible to show, as the required value is just the pathological one. Thus, for all couplings, which can be analyzed, the conformal nature is manifest.

\subsection{Numerical values of the exponent $\kappa$ and the prefactors}

The only remaining part is then to determine the actual value of $\kappa$ and the remaining coefficients. To do so, it is useful to note that the factor 
\be
\omega=\frac{C_ag^2\Gamma^2(2-\kappa)\Gamma(2\kappa)3a^2s}{48\pi^2\Gamma(4-2\kappa)\Gamma^2(1+\kappa)}
\ee
\no can be extracted from all loop integrals. As a consequence, the conditions for a solution for the four equations read
\bea
\omega&=&-\frac{(1+\kappa)(2+\kappa)}{18(-3+2\kappa)}\nn\\
\omega&=&\frac{4\kappa-2}{3(1+864 B+16F^2(2\kappa-3))}\nn\\
\omega&=&-\frac{(1+\kappa)(2+\kappa)}{14400F^2(3-8\kappa+4\kappa^2)}\nn\\
\omega&=&\frac{(2\kappa-1)(2+3\kappa+\kappa^2)}{864(2\kappa-3)(2F^2+72B+3F^2\kappa-144B\kappa+F^2\kappa)}\nn,
\eea
\no where $F=a/f$, $B=S(b/a)^2=9S(s/a)^2$. These are thus four equations for the four unknown quantities $a^2s$, $F$, $B$, and $\kappa$.

Eliminating the three constants yields a conditional equation for $\kappa$,
\be
-1=\frac{50(1-2\kappa)^2(146-381\kappa+193\kappa^2)}{(1+\kappa)(2+\kappa)(218-823\kappa+819\kappa^2)}
\ee
\no which is just a fourth-order polynomial for $\kappa$. Two solutions are complex, one is larger than one, and only one solution is between 0 and 1, and therefore provides propagators still compatible with the interpretation of a tempered distribution. The value is approximately
\be
\kappa\approx0.691354\nn
\ee
\no which is a value quiet close to the value of the corresponding exponent in the so-called scaling solution of Yang-Mills theory of about $0.59$ \cite{Zwanziger:2002ia,Lerche:2002ep,Fischer:2006vf,Fischer:2009tn,Huber:2007kc,Pawlowski:2003hq}. The equations for the other quantities are then, quoting only positive, real solutions,
\bea
F=\frac{a}{f}&=&\frac{\sqrt{2+3\kappa+\kappa^2}}{20\sqrt{2}\sqrt{2\kappa^3+5\kappa^2+\kappa-2}}\approx0.0571506\nn\\
B=9S\frac{s^2}{a^2}&=&\frac{1906-8445\kappa+11747\kappa^2-4902\kappa^3}{43200(-2+\kappa+5\kappa^2+2\kappa^3)}\approx0.00828921\nn\\
3a^2s g^2 C_a&=&\frac{2^{5-2\kappa}\pi^\frac{3}{2}(1+\kappa)(2+\kappa)\Gamma\left(\frac{3}{2}-\kappa\right)\Gamma(1+\kappa)^2}{3\Gamma(2\kappa)}\approx 110.876\nn
\eea
\no which also shows that the result is valid for any gauge coupling and gauge group. Note that there are in total four quantities still undetermined, $a$, $f$, $s$, and $S$, and only their ratios are fixed, leaving an over-all scale open. But, most importantly, all of these can have positive values, creating suitable propagators. The only remarkable feature is that the gluon and scalar have necessarily a prefactor of the same size, while the one of the fermion is about three orders smaller than that of the ghost. There is no fixed relation between the one of the ghost and the one of the scalar, so these two can once more differ substantially, which also yields possibly again a very different order of magnitude for the pre-factor of the scalar-gluon vertex.

\subsection{Summarizing the result}

This completes the solution, 
\bea
G(x)&=&ax^{-\kappa}\nn\\
Z(x)&=&3s(a)x^{2\kappa}\nn\\
F(x)&=&f(a)x^{-\kappa}\nn\\
S(x)&=&s(a)x^{2\kappa}\nn\\
\Gamma^{\phi^\ast\phi A}(x,y,z)&=&S(a)x^{\alpha_1}y^{\alpha_1}z^{6\kappa-2\alpha_1}\nn
\eea
\no where the ghost prefactor has been chosen to be the independent constant, and $\alpha_1$ is some arbitrary exponent. The three-gluon vertex is only constrained by the demand that the gluon loop cancels itself. All other vertices remain bare. The value of $\kappa$ is about 0.69, and the ratios of all remaining prefactors is fixed.

This solution is highly non-trivial, and at the same time realizes the conformal properties of the theory. Note that, like in the scaling solution of Yang-Mills theory \cite{vonSmekal:1997is,Zwanziger:2002ia,Lerche:2002ep} the ghost diverges, while the gluon vanishes. In addition, the fermions also diverge in the infrared, while the scalars also vanish, mirroring the Yang-Mills sector.

\section{The Gribov-Singer ambiguity}\label{s:gribov}

One topic which should be briefly discussed here is the question of the Gribov-Singer ambiguity. It has been conjectured for Yang-Mills theory in three and four dimensions \cite{Maas:2009se,Fischer:2008uz} that the differing ways of treating Gribov copies could reflect upon itself in the infrared properties of the correlation functions, creating a family of solutions. Such a family is indeed observed in the solutions of DSEs for Yang-Mills theory \cite{Fischer:2008uz,Boucaud:2008ji}, and some limited, but not yet conclusive, support is available from lattice calculations \cite{Maas:2009se,Maas:2011se,Sternbeck:2012mf,Maas:2013vd}. In Yang-Mills theory, the different solutions differ by the infrared behavior, characterized by the inverse ghost dressing function at zero momentum which varies continuously between 0 and some finite number of order 1. A value different from 0 is in one-to-one correspondence with an infrared finite gluon propagator.

Such a behavior would not be compatible with conformal symmetry, as the finite gluon propagator would then define a mass scale. Besides the obvious possibility that there is no Gribov-Singer ambiguity in ${\cal N}=4$ SYM, as argued in \cite{Capri:2014tta}, there are two more options. One is that since these are statements about gauge-dependent quantities, they may also break conformal symmetry, just like the Landau gauge condition already breaks supersymmetry, while all gauge-invariant quantities remain manifestly conformal (and supersymmetric).

The other option comes from the situation in two dimensions, which is somewhat different. In this case the ghost dressing function is necessarily infrared divergent \cite{Huber:2012td,Dudal:2012td,Zwanziger:2012se}. There is, however, some indications from lattice calculations \cite{Maas:2009se,Maas:2011se,Maas:2015nva} that in this case the approach to infinity may be modified, but this requires many more tests. This would yield a possibility consistent with both conformal symmetry and a non-trivial Gribov-Singer ambiguity. The latter is, however, remarkable, as also in two dimensions the argument was made that the Gribov-Singer ambiguity is essentially different from higher dimensions \cite{Dudal:2008xd}, in a very similar way as in ${\cal N}=4$ SYM \cite{Capri:2014tta}. Thus, it may well be that the situation in both cases are similar, as may be expected from the absence of dynamics in both cases.

The first case requires to find full, non-conformal solutions to the system, which is already in the Yang-Mills case a formidable problem \cite{Maas:2011se,Fischer:2008uz}. The other option will require modification of the vertices and/or inclusion of two-loop diagrams, as with the present truncation there is one and only one solution. However, already the analytic solutions for the conformal case of the diagrams appearing in three-point vertex equations are far more complicated than for the propagator equations \cite{Alkofer:2008dt}, and thus will require a more elaborate investigation. Also, an all-equations construction like in Yang-Mills theory \cite{Fischer:2006vf,Fischer:2009tn,Huber:2009wh} is at least involved, as in ${\cal N}=4$ SYM all diagrams could in general contribute, and it is not clear whether dominant diagrams can be identified to simplify the solution. Thus, both approaches are beyond the present scope.

A possible check would also be the investigation of the Gribov-Singer ambiguity using lattice methods, as such calculations for ${\cal N}=4$ SYM are possible \cite{Catterall:2007kn,Catterall:2009it}. This is under way. 

\section{Summary}\label{s:sum}

We have constructed a conformal solution for the Landau-gauge DSEs in ${\cal N}=4$ SYM in the same approximation which has been used for a large number of corresponding studies in ordinary Yang-Mills theory. Thus, this approximation conserves the relevant qualitative properties. In fact, we find consistently for all couplings accessible in our truncation a conformal behavior. The only serious problem encountered was connected with the three-gluon vertex, which in this truncation required self-cancellations to not spoil the solution.

Though this is an encouraging result, the analysis of it with respect to the Gribov-Singer ambiguity shows that there is still a lot not understood, and several avenues for future research have been outlined. Eventually, it should be possible to find a full solution of the theory using DSEs. But this will still require serious effort.\\

\no{\bf Acknowledgments}\\

We are grateful to Markus Huber, Jan Pawlowski, and Andreas Wipf for a critical reading of the manuscript and helpful comments.

\bibliographystyle{bibstyle}
\bibliography{bib}

%%%%%%%%%%%%%%%%%%%%%%%%%%%%%%%%%%%%%%%%%%%%%%%%%%%%%%%%%%%%%%%%%%%%%%%%%%%%%%%%%%%%%%%%%%%%%%%%%%

\end{document}